\documentclass[twocolumn,superscriptaddress,showpacs]{revtex4}
\usepackage{amssymb}
\usepackage{amsmath}
\usepackage{txfonts}
\usepackage{graphicx,latexsym}
\usepackage{subfigure}
\DeclareMathOperator{\Tr}{Tr} 
\DeclareMathOperator{\RIBQDD}{RIBQDD} \DeclareMathOperator{\qd}{QD}
\DeclareMathOperator{\GQD}{GQD}\DeclareMathOperator{\hsd}{HSD}

\DeclareMathOperator{\LMIMD}{LMIMD}
\DeclareMathOperator{\LEMID}{LEMID}
\DeclareMathOperator{\LEMIMD}{LEMIMD}
\DeclareMathOperator{\GMLMIMD}{GMLMIMD}
\DeclareMathOperator{\GMLEMID}{GMLEMID}
\DeclareMathOperator{\GMLEMIMD}{GMLEMIMD}
\DeclareMathOperator{\LMLMIMD}{LMLMIMD}

\DeclareMathOperator{\LMLEMID}{LMLEMID}
 \DeclareMathOperator{\ggqd}{GGQD}
\DeclareMathOperator{\gqd}{GQD} \DeclareMathOperator{\gmgqd}{GMGQD}
 \DeclareMathOperator{\LMQD}{LMQD}
\DeclareMathOperator{\LMGQD}{LMGQD}
\DeclareMathOperator{\LMVNE}{LMVNE}
\DeclareMathOperator{\cgqd}{CGQD}
\DeclareMathOperator{\ggmGQD}{GGMGQD}
\DeclareMathOperator{\ggmLMIMD}{GGMLMIMD}

\usepackage{CJK}
\usepackage{amssymb}
\usepackage{graphicx}
\usepackage{epsf}
\usepackage{dcolumn}
\usepackage{bm}
\usepackage{longtable}
\usepackage{CJK}
\usepackage{color}
\begin{document}
\begin{CJK}{GBK}{song}

\title{Multipartite Two-partite Quantum Correlation and Its Three Types of Measures}
\author{Jing-Min Zhu} \email{zhjm-6@163.com} \affiliation{College
of Optoelectronics Technology, Chengdu University of Information
Technology, Chengdu 610225, China}

\begin{abstract}
Multipartite quantum correlation (MQC) not only explains many novel
microscopic and macroscopic quantum phenomena, but also holds
promise for specific quantum technologies with superiorities. MQCs
descriptions and measures have been an open topic, due to their rich
and complex organization and structure. Here reconsidering MQC
descriptions and their practical applications in some quantum
technologies, we propose a novel description called multipartite
two-partite QC, which provides an intuitive and clear physical
picture. Specifically, we present three types of measures: one class
based on minimal entropy-like difference of local measurement
fore-and-aft multipartite two-partite density matrix such as
multipartite two-partite quantum discord (QD), another class based
on minimal trace-like geometric distance such as multipartite
two-partite Hilbert-Schmidt Distance (HSD), and a third class based
on decoherence such as multipartite two-partite Local
Measurement-Induced Minimal Decoherence (LMIMD) and Local
Eigen-Measurement-Induced Decoherence (LEMID). Their computations
required for these measures are relatively easy. All of the
advantages make them promising candidates for specific potential
applications in various quantum technologies. Finally, we employ
these three types of measures to explore the organization and
structure of some typical genuine MQCs, and analyze their relative
characteristics based on their physical implications and
mathematical structures.


\vspace{0.1in} \noindent{\bf \small{Keywords}}: Multipartite Quantum
Correlation (MQC); Multipartite Two-partite QC; Multipartite
Two-partite Quantum Discord (QD); Multipartite Two-partite
Hilbert-Schmidt Distance (HSD); Multipartite Two-partite Local
Measurement-Induced Minimal Decoherence (LMIMD); Multipartite
Two-partite Local Eigen-Measurement-Induced Decoherence (LEMID);

\end{abstract}
 \makeatletter
\makeatother

\maketitle
\section{Introduction}


Unimaginable properties of multipartite quantum correlation (MQC)
not only explain numerous novel microscopic and macroscopic quantum
phenomena, such as the Quantum Hall Effect, Bose-Einstein
condensation, and superconductivity \cite{yy,km}, but also hold
significance in specific quantum technologies with unique
functionalities and advantages \cite{h,h1,i,z,c,v,g1,f1,b,m1}, such
as quantum optical interferometry \cite{djp} and quantum lithography
\cite{mda}. The organization and structure of MQCs, due to the
inherent complexity arising from the interactions among numerous
parties, are extraordinarily rich and complex. Consequently, the
descriptions and measures of MQC have remained an open area of
research.

Over the past decade or two, significant efforts from numerous
research groups worldwide have led to the development of various
descriptions and measures for MQC. These measures include the
vector-like multiple entropy measure based on geometric mean entropy
\cite{long0}, the global multipartite discord based on relative
entropy and local Von Neumann measurements \cite{c1}, the sum of
bipartite discords obtained from successive measurement \cite{m},
the genuine MQC characterized by nonzero QC in every bipartite cut
based on relative entropy \cite{g}, the quantum dissension defined
as the difference between tripartite mutual information based on
three-variable mutual information and its generalization to quantum
setting after a single measurement \cite{i1}, the multipartite
quantum relative entropy based on local measurement
\cite{k,k1,mx,ab} and multiple quantum discord based on conditional
measurements \cite{cha}. Additionally, distance-based multiple
geometric quantum discord based on global measurements \cite{xjw}
and multiple geometric quantum discord based on conditional
measurements \cite{chen,asm} have also been proposed.

Here reconsidering MQC descriptions and their practical applications
in some quantum technologies, we propose a novel approach to
describe MQC called multipartite two-partite QC, which provides an
intuitive and clear physical MQC description. Specifically, we
present three types of measures: the first class based on minimal
entropy-like difference of local measurement fore-and-aft
multipartite two-partite density matrix such as multipartite
two-partite quantum discord (QD), the second class based on minimal
trace-like geometric distance such as multipartite two-partite
Hilbert-Schmidt Distance (HSD), and the third class based on
decoherence such as multipartite two-partite Local
Measurement-Induced Minimal Decoherence (LMIMD) and Local
Eigen-Measurement-Induced Decoherence (LEMID). The computations
required for these measures are relatively easy. Furthermore, all of
the advantages make them promising candidates for specific potential
applications with superiorities in various quantum technologies.
Subsequently, we employ these three types of measures to explore the
organization and structure of some well-known genuine MQCs, and
analyze their relative characteristics based on their physical
implications and mathematical structures.



\section{Three Types of Bipartite QC Measures}

Based on Von Neumann measurement (Orthogonal Projection Measurement
with Rank One), the existing bipartite QC measures are classified
into three categories as below.


\subsection{Bipartite Quantum Discord (QD)}

Let us begin with the first class based on minimal entropy-like
difference of local measurement fore-and-aft bipartite density
matrix, such as relative entropy of discord \cite{mk}, one-way
quantum deficit \cite{sah} and so on. Among these, a well-known and
commonly used bipartite QC measure quantum discord (QD), the minimal
difference of two kinds of classical-equivalent mutual information
over all sets of one-side orthogonal normalization complete basis
(ONCB) measurement
 ($\prod_{1}$) \cite{oh,zhang2,zhou} is described by
\begin{eqnarray}
\qd&=&S(\rho_1)+S(\rho_2)-S(\rho)\nonumber\\
&&-\max_{\prod_{1}}(S(\rho_{1\prod_{1}})+S(\rho_2)-S(\rho_{\prod_{1}}))\nonumber\\
&=&\min_{\prod_{1}}[S(\rho_1)-S(\rho)-(S(\rho_{1\prod_{1}})-S(\rho_{\prod_{1}}))],
\end{eqnarray}
where $\rho$ represents bipartite density matrix and $\rho_i
(i=1,2)$ its subsystem reduced density matrix, and $S(P)=-\Tr
P\log_{2}{P}$ ($P=\rho, \rho_1, \rho_2$) represents Von Neumann
entropy of $P$. Notably, when the bipartite quantum system is in
pure state, the QD reduces to the Von Neumann entropy of the
measurement partite reduced density matrix, given by
\begin{eqnarray}
\qd=S(\rho_1).
\end{eqnarray}
QD in fact represents the minimal mutual information loss of Von
Neumann entropy due to one-sided optimal ONCB measurement in the
context of bipartite QC.



\subsection{Bipartite Hilbert-Schmidt Distance (HSD)}

The second class of bipartite QC measures is based on the minimal
trace-like geometric distance \cite{da,ls3,fm,zhu10}. Among these
measures, one notable example is the Hilbert-Schmidt distance (HSD)
between the local measurement fore-and-aft bipartite density matrix
$\rho$ and $\rho_{\prod_{1}}$ \cite{da} described by
\begin{equation}
\hsd=\min_{\prod_1}\sqrt{\Tr[(\rho-\rho_{\prod_{1}})^2}].
\end{equation}

\subsection{Local Measurement-Induced Minimal Decoherence (LMIMD) and Its Variant Local Eigen-Measurement-Induced
Decoherence (LEMID)}

The third class of bipartite QC measures is the Local
Measurement-Induced Minimal Decoherence (LMIMD)
over all sets of the measurement partite ONCB measurement and the
other partite ONCB defined as \cite{zhu21}
\begin{eqnarray}
\LMIMD&&=\min_{\prod_{1},\prod_{2}'}\sum_{i_{1},j_{1},i_{2}',j_{2}'=1,i_{1}\neq
j_{1}}^d \mid\rho_{i_{1} i_{2}',j_{1}
j_{2}'}\mid\nonumber\\
&&=\min_{\prod_{1},\prod_{2}'}\sum_{i_1,j_1,i_2',j_2'=1,i_1\neq
j_1}^d \mid\langle i_1 i_2'|\rho|j_1 j_2'\rangle\mid.
\end{eqnarray}
For any pure state, the LMIMD reduces to Local
Eigen-Measurement-Induced Decoherence (LEMID) by considering the
measurement partite ONCB eigen-measurement ($\prod_{e1}$) and the
other partite eigen-ONCB ($\prod_{e2}$) expressed as \cite{zhu21}
\begin{eqnarray}
\LMIMD&&=\LEMID=\sum_{i_{e1},j_{e1},i_{e2},j_{e2}=1,i_{e1}\neq
j_{e1}}^d \mid\rho_{i_{e1} i_{e2},j_{e1}
j_{e2}}\mid\nonumber\\
&&=\sum_{i_{e1},j_{e1},i_{e2},j_{e2}=1,i_{e1}\neq j_{e1}}^d
\mid\langle i_{e1} i_{e2}|\rho|j_{e1} j_{e2}\rangle\mid.
\end{eqnarray}
The computation of LEMID is evidently easier than that of LMIMD.

The three types of bipartite QC measures have different physical
meanings and mathematical structures and exhibit symmetry with
respect to different partite measurements in general, only for
symmetric bipartite quantum states.

\section{Multipartite Two-partite QC and Its Three Types of Measures}
\subsection{Multipartite Bipartite QC, and Its Practical Physical Motivation and Mathematical Definition}
On one hand, in the context of MQC, whether in quantum technology or
in multipartite system, its two subsystem QC involved 
existing does not trace the other part in consideration of reality.
On the other hand, in any MQC system, all the information is
contained within its wave function or density matrix; when a
subsystem is obtained by partial tracing the other subsystem
freedom, some information is inevitably lost. Therefore, the concept
of multipartite two-partite QC is presented to address this issue,
which holds certain importance and practical significance.

To elaborate further, let's consider an \emph{N}-partite quantum
system. In the \emph{n}-partite reduced density matrix $\rho_n$
(where $3\leq n \leq N$), the density matrix of the $i_1$-th and
$i_2$-th parties, denoted as the \emph{n}-partite two-partite
($i_1$,$i_2$) density matrix, only preserves the information of the
$i_1$-th and $i_2$-th parties from $\rho_n$. This particular density
matrix is represented as $\rho(i_1,i_2)_n$. The n-partite
two-partite ($i_1$,$i_2$) QC is then described by
QC($\rho(i_1,i_2)_n$). Building upon the existing three types of
bipartite QC measures, a detailed explanation of the three types of
multipartite two-partite QC measures is provided below.

\subsection{Three Types of Multipartite Two-partite QC Measures}
\subsubsection{Multipartite Two-partite QD}

Let's start by discussing the first class based on minimal
entropy-like difference of local measurement fore-and-aft
multipartite two-partite density matrix, such as multipartite
two-partite QD. Specifically, for any \emph{N}-partite quantum
system, the \emph{n}-partite two-partite ($i_1$,$i_2$) QD is
described as
\begin{eqnarray}
&&\qd(\rho(i_1,i_2)_n)=S(\rho_{i_1}(\rho(i_1,i_2)_n))+S(\rho_{i_2}(\rho(i_1,i_2)_n))\nonumber\\&&-S(\rho(i_1,i_2)_n)
-\max_{\prod_{{i_1}}}[S(\rho_{{i_1}\prod_{i_1}}(\rho(i_1,i_2)_n))\nonumber\\&&+S(\rho_{i_2}(\rho(i_1,i_2)_n))-S({\rho(i_1,i_2)_n}_{\prod_{i_1}})],
\end{eqnarray}
where $\rho_{i_1}(\rho(i_1,i_2)_n)$ ($\rho_{i_2}(\rho(i_1,i_2)_n)$)
represents the $i_1$-th ($i_2$-th) partite reduced density matrix in
the reduced density matrix $\rho_n$. In the case of a symmetric
state, the \emph{n}-partite two-partite QD can be simplified to
\begin{eqnarray}
&&\qd(\rho(2)_n)=\qd(\rho(i_1,i_2)_n).
\end{eqnarray}
When the \emph{n}-partite two-partite ($i_1$,$i_2$) quantum system
is in pure state, the \emph{n}-partite two-partite ($i_1$,$i_2$) QD
is reduced to the Von Neumann entropy of the reduced density matrix
of the measurement partite in the density matrix $\rho(i_1,i_2)_n$
\begin{eqnarray}
\qd(\rho(i_1,i_2)_n)=S(\rho_{i_1}(\rho(i_1,i_2)_n)).
\end{eqnarray}

\subsubsection{Multipartite Two-partite HSD}

The second class of multipartite two-partite QC measures involves
minimal trace-like geometric distance about local measurement
fore-and-aft system density matrix, such as multipartite two-partite
HSD. Concretely for any \emph{N}-partite quantum system,
\emph{n}-partite two-partite ($i_1$,$i_2$) HSD is described by
\begin{equation}
\hsd(\rho(i_1,i_2)_n)=\min_{\prod_{i_1}}
\sqrt{\Tr[(\rho(i_1,i_2)_n-{\rho(i_1,i_2)_n}_{\prod_{i_1}})^2}].
\end{equation}
For symmetric states, \emph{n}-partite two-partite HSD reduces to
\begin{equation}
\hsd(\rho(2)_n)=\hsd(\rho(i_1,i_2)_n).
\end{equation}

\subsubsection{Multipartite Two-partite LMIMD and Its Variant
Multipartite Two-partite LEMID}

The third class of multipartite two-partite QC measures is
multipartite two-partite LMIMD
over all sets of the partite measurement ONCB measurement basis and
the other partite ONCB. Concretely for any \emph{N}-partite quantum
system, the n-partite two-partite ($i_1$,$i_2$) LMIMD is described
by
\begin{eqnarray}
&&\LMIMD(\rho(i_1,i_2)_n)\nonumber\\&&=\min_{\prod_{i_1},\prod_{i_2}'}\sum_{i_{1},j_{1},i_{2}',j_{2}'=1,i_{1}\neq
j_{1}}^d \mid{\rho(i_1,i_2)_n}_{i_{1} i_{2}',j_{1}
j_{2}'}\mid\nonumber\\
&&=\min_{\prod_{i_1},\prod_{i_2}'}\sum_{i_1,j_1,i_2',j_2'=1,i_1\neq
j_1}^d \mid\langle i_1 i_2'|{\rho(i_1,i_2)_n}|j_1 j_2'\rangle\mid.
\end{eqnarray}
For a symmetric state, the \emph{n}-partite two-partite LMIMD
reduces to
\begin{eqnarray}
&&\LMIMD(\rho(2)_n)=\LMIMD(\rho(i_1,i_2)_n).
\end{eqnarray}
For a pure symmetric \emph{n}-partite two-partite state, the
\emph{n}-partite two-partite LMIMD reduces to its LEMID
\begin{eqnarray}
&&\LMIMD(\rho(2)_n)=\LEMID(\rho(2)_n)=\LEMID(\rho(i_1,i_2)_n)\nonumber\\&&=\sum_{i_{e1},j_{e1},i_{e2},j_{e2}=1,i_{e1}\neq
j_{e1}}^d \mid\rho_{i_{e1} i_{e2},j_{e1}
j_{e2}}\mid\nonumber\\
&&=\sum_{i_{e1},j_{e1},i_{e2},j_{e2}=1,i_{e1}\neq j_{e1}}^d
\mid\langle i_{e1} i_{e2}|\rho|j_{e1} j_{e2}\rangle\mid.
\end{eqnarray}
Obviously, computing the multipartite two-partite LEMID is easier
than computing the multipartite two-partite LMIMD.

In general, only for symmetric multipartite quantum states, the
specified three types of multipartite two-partite QC measures
exhibit symmetry with respect to different partite measurements. It
is evident that the physical interpretation of these multipartite
two-partite QC measures is intuitive and clear, and the computation
of these three measures is relatively simple. These facts show that
the realization of their specific potential applications with
superiorities would be relatively easy. Then we employ these three
types of measures to investigate the organization and structure of
some well-known genuine MQCs, and analyze their relative
characteristics based on their physical connotation and mathematical
structure in details as below.



\section{Applications, Advantages and Disadvantages of Three Types of Multipartite two-partite QC Measures}

First we consider the symmetric pure \emph{N}-partite GHZ-like
(\emph{N}-GHZ-like) state
\begin{equation}
|\Phi\rangle=\cos\alpha|11\cdots1\rangle +
\sin\alpha|00\cdots0\rangle,
\end{equation}
with $0\leqslant\alpha\leqslant\pi$. Obviously, all of the reduced
density matrices are the same, and hence any two-partite $(i_1,i_2)$
density matrix in all of the reduced density matrices is also the
same
\begin{eqnarray}
&&\rho(i_1,i_2)_{N-1}=\rho(i_1,i_2)_{N-2}=\cdots=\rho(i_1,i_2)\nonumber\\
&&=\cos^2\alpha|11\rangle\langle 11|+
\sin^2\alpha|00\rangle\langle00|,
\end{eqnarray}
while only in the specified \emph{N}-GHZ-like state, any two-partite
state, namely any \emph{N}-partite two-partite ($i_1$,$i_2$) state
is
\begin{equation}
|\Phi(2)_N\rangle=|\Phi(i_1,i_2)_N\rangle=\cos\alpha|11\rangle+\sin\alpha|00\rangle.
\end{equation}
These facts show that the specified \emph{N}-GHZ-like state only has
global or \emph{N}-partite two-partite QC, while lacking local
multipartite two-partite QC or local MQC. Specifically, the global
or \emph{N}-partite two-partite QD, the global or \emph{N}-partite
two-partite HSD and the global or \emph{N}-partite two-partite LMIMD
are respectively as below
\begin{eqnarray}
&&\qd(\rho(2)_N)=\qd(\rho(i_1,i_2)_N)\nonumber\\&&=-(\cos^2\alpha\log_{2}{\cos^2\alpha}+\sin^2\alpha\log_{2}{\sin^2\alpha}),
\end{eqnarray}
\begin{eqnarray}
&&\hsd(\rho(2)_N)=\hsd(\rho(i_1,i_2)_N)\nonumber\\&&=\frac 1
2\sqrt{\frac 1 2 (3-\cos4\alpha-2\cos^22\alpha)}
\end{eqnarray}
and
\begin{eqnarray}
&&\LMIMD(\rho(2)_N)=\LMIMD(\rho(i_1,i_2)_N)\nonumber\\&&\equiv\LEMID(\rho(i_1,i_2)_N)=\mid\sin2\alpha\mid
\end{eqnarray}
as shown in Fig.1 which shows that their monotonicity is consistent,
with their curves only coinciding at a few points. This discrepancy arises from the differences in their physical connotation and mathematical structure. 
Specifically, when $\alpha=\frac\pi4, \frac{3\pi}4$, any
\emph{N}-partite two-partite state that achieves maximal QC will
have maximal values for all measures-
$\qd(\rho(i_1,i_2)_N)=\LMIMD(\rho(i_1,i_2)_N)=\LEMID(\rho(i_1,i_2)_N)=1$,
while $\hsd(\rho(i_1,i_2)_N)=\frac 1 {\sqrt2}$ which does not
reflect the degree of \emph{N}-partite two-partite QC here. On the
other hand, when $\alpha=0, \frac{\pi}2, \pi$, any \emph{N}-partite
two-partite state having no QC,
$\qd(\rho(i_1,i_2)_N)=\hsd(\rho(i_1,i_2)_N)=\LMIMD(\rho(i_1,i_2)_N)=\LEMID(\rho(i_1,i_2)_N)=0$.
It is evident that, in the case of the specified \emph{N}-GHZ-like
state, \emph{N}-partite two-partite HSD may not reflect the degree
of \emph{N}-partite two-partite QC, while \emph{N}-partite
two-partite QD, and \emph{N}-partite two-partite LMIMD or LEMID
serve as reliable measures for assessing MQC. These three types of
measures indeed can reflect the presence or absence of
\emph{N}-partite two-partite QC here.

\begin{figure}
\setlength{\abovecaptionskip}{0.1pt}%
\setlength{\belowcaptionskip}{0.1pt}%
\begin{center}
\includegraphics[width=0.45\textwidth]{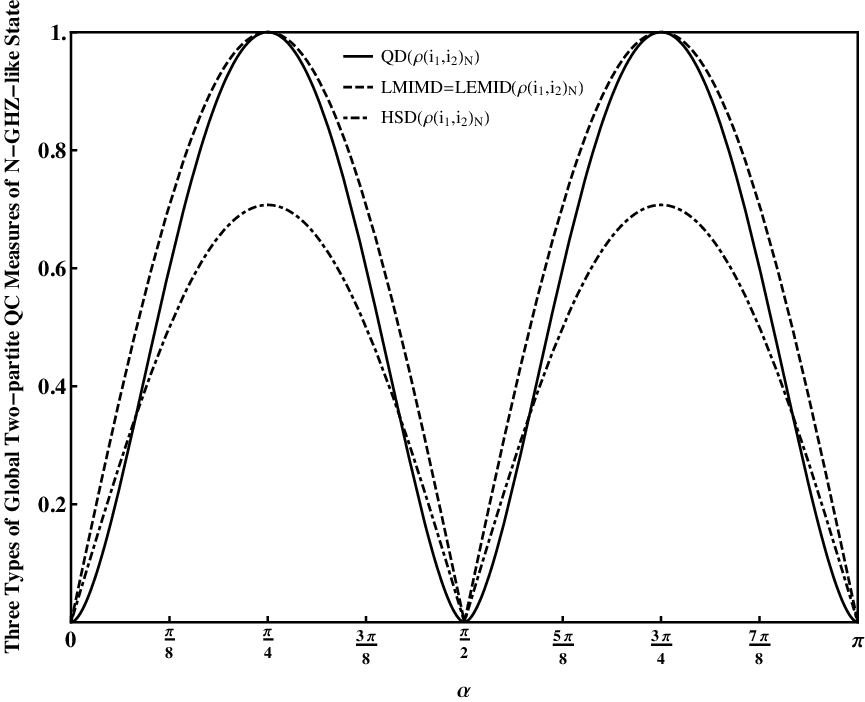}
\renewcommand{\figurename}{Fig.}
\caption{Three types of global two-partite QC measures of N-GHZ-like
state varying with $\alpha$. When $\alpha=\frac\pi4, \frac{3\pi}4$,
$\qd(\rho(i_1,i_2)_N)=\LMIMD(\rho(i_1,i_2)_N)=\LEMID(\rho(i_1,i_2)_N)=1$
, while $\hsd(\rho(i_1,i_2)_N)=\frac 1 {\sqrt2}$. When $\alpha=0,
\frac{\pi}2, \pi$,
$\qd(\rho(i_1,i_2)_N)=\hsd(\rho(i_1,i_2)_N)=\LMIMD(\rho(i_1,i_2)_N)=\LEMID(\rho(i_1,i_2)_N)=0$.}
\end{center}
\end{figure}

We next consider the symmetric pure W state
\begin{equation}
|\Phi\rangle=\frac 1 {\sqrt 3} |100\rangle + \frac 1 {\sqrt 3}
|010\rangle + \frac 1 {\sqrt 3} |001\rangle.
\end{equation}
Any two-partite $(i_1,i_2)$ pure state in the specified symmetric
pure W state, namely any tripartite two-partite $(i_1,i_2)$ state is
\begin{equation}
|\Phi(i_1,i_2)\rangle_3=\frac 1 {\sqrt 3} |10+01\rangle +\frac 1
{\sqrt 3} |00\rangle,
\end{equation}
with its single-site Eigensystem $\{\{\frac 1 6(3+\sqrt 5), \frac 1
6(3-\sqrt 5)\};\{\{0.525731\}, \{0.850651\}\};\{\{-0.850651\},
\{0.525731\}\}\}$. 
Its three types of QC measures, any global or tripartite two-partite
QD, HSD, and LMIMD are respectively as below
\begin{eqnarray}
\qd(\rho(i_1,i_2)_3)=0.550048,
\end{eqnarray}
\begin{equation}
\hsd(\rho(i_1,i_2)_3)=0.471405
\end{equation}
and
\begin{eqnarray}
\LMIMD(\rho(i_1,i_2)_3)\equiv\LEMID(\rho(i_1,i_2)_3)=\frac 2 3.
\end{eqnarray}
For pure symmetric states, the measured partite optimal ONCB for the
three types of tripartite two-partite QC measures are determined by
the eigenvectors of its reduced density matrix. Additionally, for
the global or tripartite two-partite LMIMD measure, the other
partite ONCB also correspond to its same eigenvectors. These three
types of measures values exhibit differences, which arise from their
different physical meanings and mathematical structures.

Any two-partite $(i_1,i_2)$ symmetric reduced density matrix is
\begin{eqnarray}
&&\rho(i_1,i_2) =\frac 1 3 |10+01\rangle\langle10+01| + \frac 1
3|00\rangle\langle00|,
\end{eqnarray}
with its single-site Eigensystem $\{\{\frac 1 3, \frac 2 3\};
\{\{1\}, \{0\}\};\{\{0\}, \{1\}\}\}$. If the first part represents
QC term with a probability of $\frac 2 3$, while the second part has
no QC with a probability of $\frac 1 3$, hence the two-partite QC is
not larger than $\frac 2 3$.
Its three types of QC measures, namely any two-partite QD, HSD and
LMIMD are respectively as below
\begin{eqnarray}
\qd(i_1,i_2)=0.085622
\end{eqnarray}
with the measured partite ONCB measurement basis $\{\{\{0.525731\},
\{0.850651\}\};\{\{-0.850651\}, \{0.525731\}\}\}$,
\begin{equation}
\hsd(\rho(i_1,i_2))=0.408248
\end{equation}
with the measured partite ONCB measurement basis $\{\{\{\frac {\sqrt
2} 2\}, \{\frac {\sqrt 2} 2\}\};\{\{-\frac {\sqrt 2} 2\},
\{\frac {\sqrt 2} 2\}\}\}$, 
and
\begin{eqnarray}
&&\LMIMD(\rho(i_1,i_2))=\frac 2 3
\end{eqnarray}
with the same measured partite basis as the other partite ONCB
$\{\{\{1\}, \{0\}\};\{\{0\},
\{1\}\}\}$. 
These three types of measures values exhibit significant
differences, which arise from their different physical meanings and
mathematical structures, and completely different measurement bases. 



\section{Conclusion}
In this paper, we undertake a fundamental reconsideration of the
descriptions and some practical applications of MQC in 
quantum technologies, and propose a novel approach to describe MQC,
namely multipartite two-partite QC, which offers an intuitive and
clear physical interpretation. Specifically, we present three types
of measures: the first class based on minimal entropy-like
difference of local measurement fore-and-aft multipartite
two-partite density matrix such as multipartite two-partite QD, the
second class based on minimal trace-like geometric distance such as
multipartite two-partite HSD, and the third class based on
decoherence such as multipartite two-partite LMIMD and LEMID. These
measures offer relatively simple computational methods for assessing
MQC. Furthermore, their advantages make them promising candidates
for specific potential applications with superiorities in various
quantum technologies. In general, only for symmetric multipartite
quantum states, they exhibit symmetry with respect to different
partite measurements. Using these three types of measures, we
explore the organization and structure of some typical genuine
MQCs. 
The analysis reveals that, in general, these three types of
measurement values exhibit differences, and there may even be
significant differences. These differences arise from their distinct
physical meanings and mathematical structures. 
In future studies, reduction-induced multipartite two-partite QC
decrease and its three types of measures deserve to be extended to
describe and measure MQC.

\end{CJK}
\end{document}